\documentclass[a4paper]{article}
\usepackage{amsmath,amssymb,amsfonts,amsthm} 
\usepackage{mathrsfs} 
\usepackage{enumerate} 
\usepackage{cases}
\usepackage{float} 
\usepackage{makecell}
\usepackage{indentfirst}
\usepackage{booktabs}
\usepackage{graphicx} 
\allowdisplaybreaks[4] 
\usepackage{wrapfig}  
\usepackage{url} 
\usepackage[usenames]{color} 
\usepackage{setspace} 
\def\b1{\mbox{\boldmath $1$}} 

\makeatletter
\newcommand{\Biggg}{\bBigg@{3.5}} 
\makeatother
\theoremstyle{plain} 
\newtheorem{theorem}{\bf Theorem}[section] 
\newtheorem{corollary}{\bf Corollary}[section] 
\newtheorem{proposition}{\bf Proposition}[section] 
\newtheorem{definition}{\bf Definition}[section] 
\newtheorem{remark}{\bf Remark}[section]  
\newtheorem{example}{\bf Example}[section] 

\makeatother
\allowdisplaybreaks 

\usepackage[numbers,sort&compress]{natbib} 
\usepackage[bookmarksnumbered, bookmarksopen,colorlinks,citecolor=blue,linkcolor=blue]{hyperref} 

\begin{document}
\onehalfspacing 
\date{}

\title{A novel k-generation propagation model for cyber risk and its application to cyber insurance }
\author{  Na Ren, 
	\quad Xin Zhang  \thanks{Corresponding author: Xin Zhang , email:x.zhang.seu@gmail.com }
	\quad\\
	School of Mathematics\\
	Southeast University\\
	Nanjing, 211189, P.R.China\\
}
\date{}
\maketitle \noindent {\normalsize{\bf Abstract:}
The frequent occurrence of cyber risks and their serious economic consequences have created a growth market for cyber insurance. The calculation of aggregate losses, an essential step in insurance pricing, has attracted considerable attention in recent years. This research develops a path-based k-generation risk contagion model in a tree-shaped network structure that incorporates the impact of the origin contagion location and the heterogeneity of security levels on contagion probability and local loss, distinguishing it from most existing models.
Furthermore, we discuss the properties of k-generation risk contagion among multi-paths using the concept of d-separation in Bayesian network (BN), and derive explicit expressions for the mean and variance of local loss on a single path. By combining these results, we compute the mean and variance values for aggregate loss across the entire network until time $t$, which is crucial for accurate cyber insurance pricing. Finally, through numerical calculations and relevant probability properties, we have obtained several findings that are valuable to risk managers and insurers.
}   

  
\vskip 0.3cm

\section{Introduction}
While embracing the convenience brought by the Internet, all sectors are facing unprecedented risk crises, such as information leakage, computer creep attacks, etc., which result in significant economic losses.
Cyber risks constitute a severe threat to companies worldwide. \cite{eling_cyber_2020,marked_2022} reported that
the huge financial impact of various cyber incidents has recently intensified due to their increasing occurrence rate, with business interruption (BI) and information loss having the highest monetary consequences.
The estimated cost for data breach incidents, as stated in \cite{NIST2}, could reach several million USD on average, while the annual global costs of cyber risk are approximately one hundred billion USD. 
The largest recorded cyber claim reached a staggering US \$80 million, as revealed in NetDiligence's 2019 report\cite{nist}. Additionally, the cost per-record exceeded an astonishing US \$1.5 million. Moreover, McAfee's research conducted in 2014\cite{McAfee} and insights provided by the World Economics Forum both underscore the significant financial implications associated with cyber risks.
Addressing these risks has become a prominent concern for both industries and scholars alike. As \cite{gordon_framework_2003} demonstrated, in addition to enhancing security technologies at various levels to mitigate such threats, developing effective strategies for managing cyber security through insurance is crucial for efficiently transferring risks while minimizing potential losses.
The development of cyber insurance is fraught with challenges for insurers and risk managers who seek to offer appropriate insurance products and derive profits from them. However, as reported in \cite{cyber_actuarial,Carfora2018}, the complexity of the cyber risk landscape, limited availability of historical loss data pertaining to risk events, diverse policy regulations, and information asymmetry between the two parties involved in insurance transactions all contribute to the nascent stage of cyber security insurance development.

From an actuarial perspective, the accurate assessment of risk loss plays a fundamental role in both risk mitigation by risk managers and premium pricing for insurers \cite{cyberloss2016,ELING20191109}. \cite{eling_what_2016} pointed that the intricate interdependence of cyber risks across sectors and businesses poses challenges to the aforementioned task, necessitating urgent need to model the dependence of cyber risks. Studies \cite{eling_goodness_prc_2017,xu_modeling_2018,Bessymulti,awiszus_modeling_2023} proposed various statistical approaches to model the dependence of cyber risks. Copula methods \cite{xu_modeling_2018,Eling_copula_2018} have emerged as commonly used models for capturing non-linear dependencies in cyber risk. However, it is evident that the limited availability of historical data on cyber risk claims hampers the application of more advanced statistical models in understanding the mechanisms underlying cyber risk dependence. Furthermore, our focus extends beyond solely modeling the dependence of cyber risks; we also aim to explore the contagion effect resulting from these interdependencies among risks. Recently, probabilistic approaches have gained significant attention in modeling this contagion effect due to their enhanced interpretability compared to traditional statistical models.

Network topology is an effective methods for describing the interdependence between risk entities, and has been widely used in modeling financial risk dependence and contagion modeling\cite{bo_systemic_2015,kanno_network_2016,amini_systemic_2020}.
 Existing studies \cite{van_virus_2009,acemoglu_network_2016,xu_hack_2019,fahrenwaldt_pricing_2018,hillairet_contagion_2022} on this issues primarily combine network topology structures with a susceptible-infected-susceptible(SIS) epidemic spreading model, in which each node in the network corresponds to a single risk arrival process and loss process, while cyber infections are modeled using a susceptible-infected-susceptible process.
Those methods comprehensively 
capture the state transition and aggregate loss of risks within the network structure. However, high-dimensional calculations pose challenges that can be addressed through relevant approximation methods such as mean-field or simulation approaches. 
The pond percolation model, proposed by \cite{1957Percolation} and widely applied in various fields related to complex networks, is another commonly used method for contagion in the network. Therefore, it is a natural choice to apply the bond percolation model for modeling cyber risk contagion. Building upon the bond percolation model on network topology,  \cite{jevtic_dynamic_2020} developed a dynamic structural aggregate loss model specifically designed for small and medium-sized enterprises, in which each arrival of attack equips a stochastic tree network structure.
The local loss in the network caused by an origin contagion is computed, and the explicit mean and variance of aggregate loss are derived using the classical collective risk model framework.
\cite{Lanchier_2023} utilized this framework to investigated the cyber risk of a client-server network system  characterized as a random star topology, as well as a prototypical hospital system considered as a mixed network \cite{da_evaluating_2024} proposed a risk contagion model on two hybrid network topologies employing the bond percolation model. 

To quantify the cyber contagion on the network structure, as a special case of bond percolation contagion model, \cite{Lhop_2011,laszka_assessment_2018,da_multivariate_2021} proposed one-hop and multi-hop risk contagion models to capture the depth of risk contagion.
\cite{Lhop_2011} proposed L-hop percolation on networks by considering that a node can be deleted (or failed) because it is chosen or because it is within some L-hop distance of a chosen node.
\cite{da_multivariate_2021} proved that the contagion states of nodes exhibit positively associated properties for any network structure based on the k-hop model. However, most existing work attributes risk contagion solely to interconnection between nodes in the network while ignoring the impact of security levels of nodes and risk size on risk contagion.
To our knowledge, the external attack probability and contagion probability are always taken as a constant $p$ in the existing studies; comparisons between security levels and loss sizes are not considered when calculating probability $p$.
This is exactly the topic we aim to address in the present work. 

 Similar to the network setting in study \cite{jevtic_dynamic_2020}, the tree-shaped network graph is employed to gain a comprehensive understanding of the proposed risk contagion.
 This topological structure serves as a fundamental component for constructing more intricate network structures and is commonly used in military units, government units, and other organizations with strict hierarchical boundaries and clear levels. 
 To capture the influence of node heterogeneity across different layers in tree-shaped structures on risk contagion, we assume varying levels of safety (risk load levels) for nodes at different layers. Moreover, unlike the undirected graph in existing work\cite{jevtic_dynamic_2020}, the directed tree-shaped network is used to capture the risk propagation from high-security layers to low-security ones. 
 Our risk contagion model introduces a path-based k-generation risk propagation mechanism wherein the contagion initiates from a compromised origin node due to an external risk attack and spreads to its k-generation descendants. In essence, our k-generation risk contagion mechanism extends the existing k-hop contagion model at the probability distribution level. Additionally, the k-generation contagion probability is characterized by a multivariate joint probability distribution. To alleviate computational complexity associated with calculating joint probabilities, we leverage d-separation concept on directed acyclic graphs (DAGs)\cite{Graphical2009} to transform joint probabilities into products of conditional probabilities under certain conditions are met. Compared with prior studies, our work exhibits several noteworthy contributions:
\begin{enumerate}
    \item A variant of k-hop risk propagation model based on the tree-shaped network is proposed, in which the probability of k-generation risk contagion is defined as a product of conditional probabilities. 
    \item In addition, incorporating the security levels of node, network branch size, etc. risk factors into the risk propagation model to quantify the impact on risk propagation, which has been less mentioned in existing work.
    \item The mathematical framework of the aggregate loss based on the proposed k-generation risk contagion model is developed, and the numerical analysis of cyber insurance pricing is conducted.
    \end{enumerate}
    
Under the proposed k-generation risk contagion model, we calculate the probability properties of local loss which caused by an origin contagion and derive the explicit mean and variance of aggregate loss. To get a better understanding of the proposed model, we conduct a numerical calculation to analyze the impact of parameters on the mean and variance of aggregate loss. Finally, the experiment of an application to cyber insurance pricing is conducted and several finding are concluded.
The rest of this paper is organized as follows:
in section \ref{sec2}, the mathematical framework of aggregate loss is proposed. The path-based k-generation risk propagation model on the tree-shaped network structure is developed in Section \ref{sec3}. In Section \ref{sec4}, we conduct numerical calculations and some conclusions are obtained, and in Section \ref{sec5}
concludes the paper.
\section{Natations and model description}\label{sec2}
In this section, we present a mathematical framework for an aggregate loss model on a tree-shaped network structure based on the proposed k-generation risk propagation model. For convenience, we first give some representations that will be discussed later in this work.
\subsection{Notations}

\begin{tabular}{cc}
\toprule 
$t$ & the time horizon \\
$R$ & the radius of the tree-shaped network\\
$T_{R}^{i}$ & the stochastic tree-shaped network that corresponding to the i-th external risk \\
$\rho$ & the size of descendants for the tree-shaped network structure\\
$\mu$ & a constant intensity of homogeneous Poisson process \\
$\beta_{k}$ & \makecell{the adjust coefficient of rise size for the k-generation \\ risk propagation}\\
$X_{i}$ & the external risk size of i-th risk arrival \\
$X_{ki}$ & the risk size at which the k-hop risk propagation is arrivals \\
$I_{r}^{(k)}$ & the state of the event $\{\beta_{k}X>c_{r+k}\}$\\
 $\bar{I}_{r}^{(k)}$ & the state of the k-generation risk propagation along a single path  \\
 $Z_{r}^{(k)}$ & the loss on single path caused by k-generation risk propagation\\
$U_{r}^{(k)}$ & the random number of paths at which the k-generation risk contagion occurs \\
$S_{r}^{(k)}$ & the local aggregate loss that
corresponds the number $U_{r}^{(k)}$\\
$L_{rt}^{(k)}$ & \makecell{the aggregate loss caused by k-generation risk contagion\\ on the network until the time $t$}\\
\bottomrule 
\end{tabular}

\subsection{Mathematical framework of aggregate loss model} \label{collective}
In this subsection, we develop a mathematical framework to model aggregate loss $(L_{t})$ from continuous time perspective. Although many results have provided calculations for aggregate loss from the single-periods cases\cite{sarabia_aggregation_2018,zhang_dynamic_2022}, it is meaningful to study the aggregate loss generated over multiple periods in continuous time to better reflect the dynamic changes of aggregate loss over time, especially for research in cyber cyber insurance.
Our aggregate loss process $L_{t}$ is essentially a variant of the classic aggregate risk model tailored to the characteristics of cyber security risks.
The aggregate loss process is a stochastic process that is comprised of a Poisson process representing the outside risk occurrences, a tree-shaped network denoting the interconnectedness of the individuals within the system, and a cyber risk contagion dynamics model. More precisely, the aggregate loss process can be developed using the following components:
\begin{enumerate}
    \item The arrival of external risk attack $\{(T_{1},X_{1},T_{R}^{1}), (T_{2},X_{2},T_{R}^{2}),\dots\}$
 follows a marked homogeneous Poisson process (MHPP) with a constant intensity $\mu$\cite{2003Daley}. 
 The risk magnitudes $\{X_{i},i=1,2,\dots\}$ are mutually independent and follow a probability distribution with density function $f_{X}(x)$. We assume that the loss magnitudes $X_{i}$ is decreasing along the depth of risk contagion, more precisely, denote $X_{i}^{(k)}=\beta_{k}X_{i}, \beta_{k}\in(0,1)$ is the size of risk that corresponding to the depth $k$ of risk contagion.
   \item For each external risk arrival time $T_{i}$, there exists a tree-shaped network denoted by $T_{R}^{i}=(\mathbf{V}^{i},\mathbf{E}^{i})$ with the radius $R$. 
   Assume that the tree-shaped networks generated at each external risk arrival time are denoted as
   \begin{equation*}
      T_{R}^{1},T_{R}^{2},\dots,T_{R}^{i},\dots,
   \end{equation*}
   which are independently and identically distributed.
   \item Vector $c=(c_{0},c_{1},c_{2},\dots,c_{R})^{T}$ denote the risk loading level (security level) of nodes that located at a distance r from the root. It is assumed that for every tree-shaped network $T_{R}^{i}$, all nodes at a distance of $r$ from the root have the same risk load level(security level), so we omit the superscript $i$.
   \item  Risk contagion mechanism always assumes that the risk propagation occurs from each origin contagion to its offspring nodes, that is, it is a kind of directed risk contagion.
\end{enumerate}
Compared with the existing works, a marked Poisson process is used and the external loss size is considered. 
Denote the random variable $L_{t}$ as the aggregate loss caused by risk contagion on the entire network until time $t$. From the collective risk framework, 
\begin{equation}\label{loss}
L_{t}=\sum_{i=1}^{N_{t}}S_{i},
\end{equation}
where $S_{i}$ represents the local loss caused by the i-th external risk attack, and in next context we can see that the $S_{i}$ is dependent with the depth of risk propagation $k$ and the location $r$ of the original compromised node.
The formula \eqref{loss} fully describes the aggregate loss model. The framework could be generalised by any network, we mainly employ the tree-based network structure to get the explicit analytical result.
 Note that the network structure, external risk arrivals, and the risk contagion are independent and identically distributed for each risk incident. To get the moment function of aggregate loss, we have 
\begin{equation*}
\begin{aligned}
\mathbb{E}[L_{t}\vert N_{t}=n]&=n\mathbb{E}[S],\\
\mathbb{V}ar[L_{t}\vert N_{t}=n]=&
n\mathbb{V}ar[S].
\end{aligned}
\end{equation*}
By the condition expectation formula
\begin{equation}\label{form2}
    \begin{aligned}
\mathbb{E}[L_{t}]&=\mathbb{E}[\mathbb{E}(L_{t}\vert N_{t})]=\mathbb{E}[N_{t}\mathbb{E}(S)]=\mathbb{E}[N_{t}]\mathbb{E}[S],\\
\mathbb{V}ar[L_{t}]&=\mathbb{E}[\mathbb{V}ar[L_{t}\vert N_{t}]+\mathbb{V}ar[\mathbb{E}[L_{t}\vert N_{t}]]\\
&=\mathbb{E}[N_{t}\mathbb{V}ar[S]]+\mathbb{V}ar[N_{t}\mathbb{E}[S]]\\
&=\mathbb{E}[N_{t}]\mathbb{V}ar[S]+\mathbb{V}ar[N_{t}](\mathbb{E}[S])^{2},
    \end{aligned}
\end{equation}
formula \eqref{form2} shows that the mean and variance of the aggregate loss until time $t$ can be computed based on the the $\mathbb{E}[S]$ and $\mathbb{V}ar[S]$. In the next subsection, we focus on the calculation of $\mathbb{E}[S]$ and $\mathbb{V}ar[S]$ based on the proposed risk propagation model.

\section{ Path-based k-generation risk contagion model}\label{sec3}
\begin{figure}[H]
 	\centering
 	\begin{tabular}{cc}
 		\includegraphics[width=0.48\linewidth]{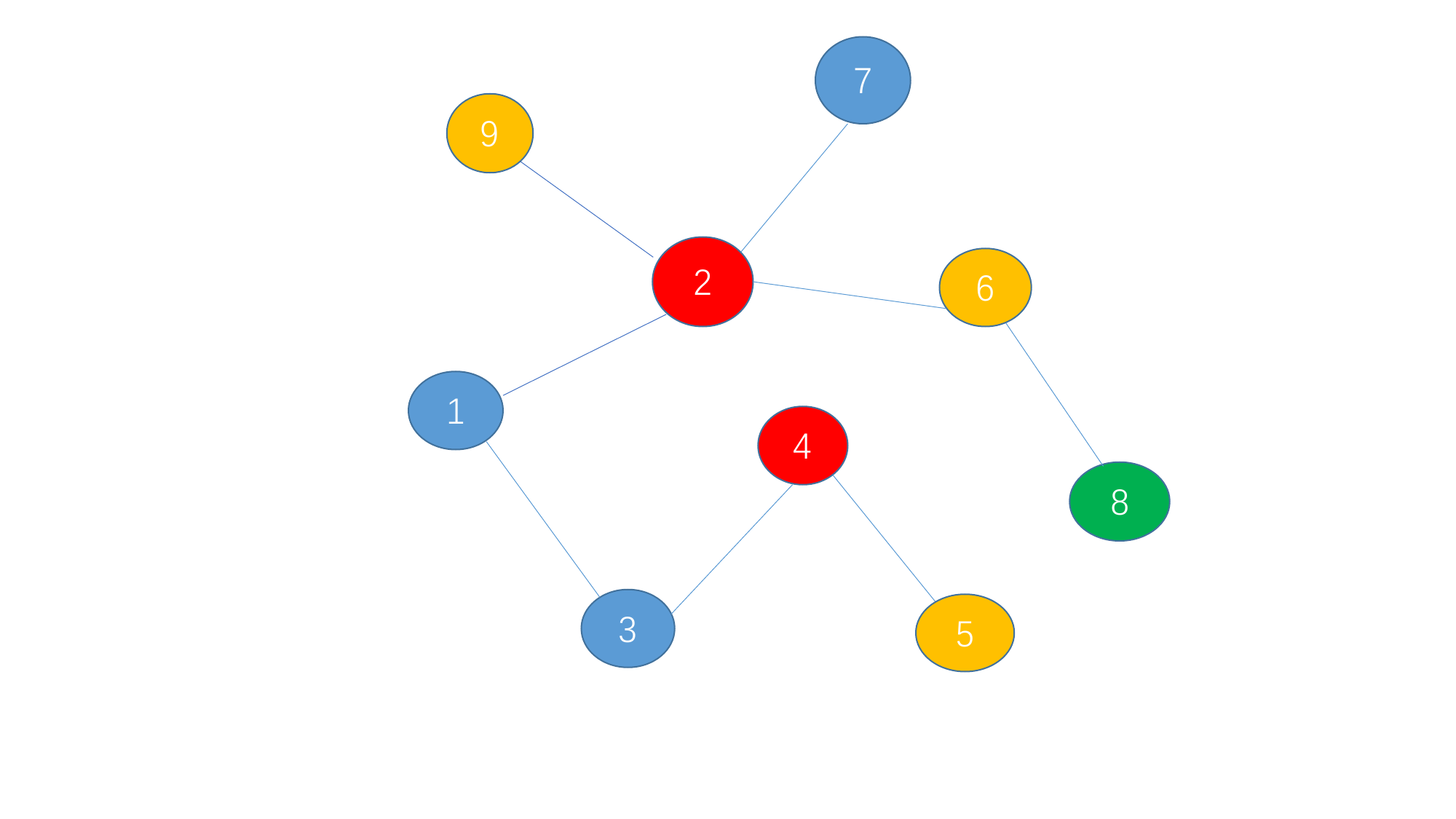}& 
 		\includegraphics[width=0.48\linewidth]{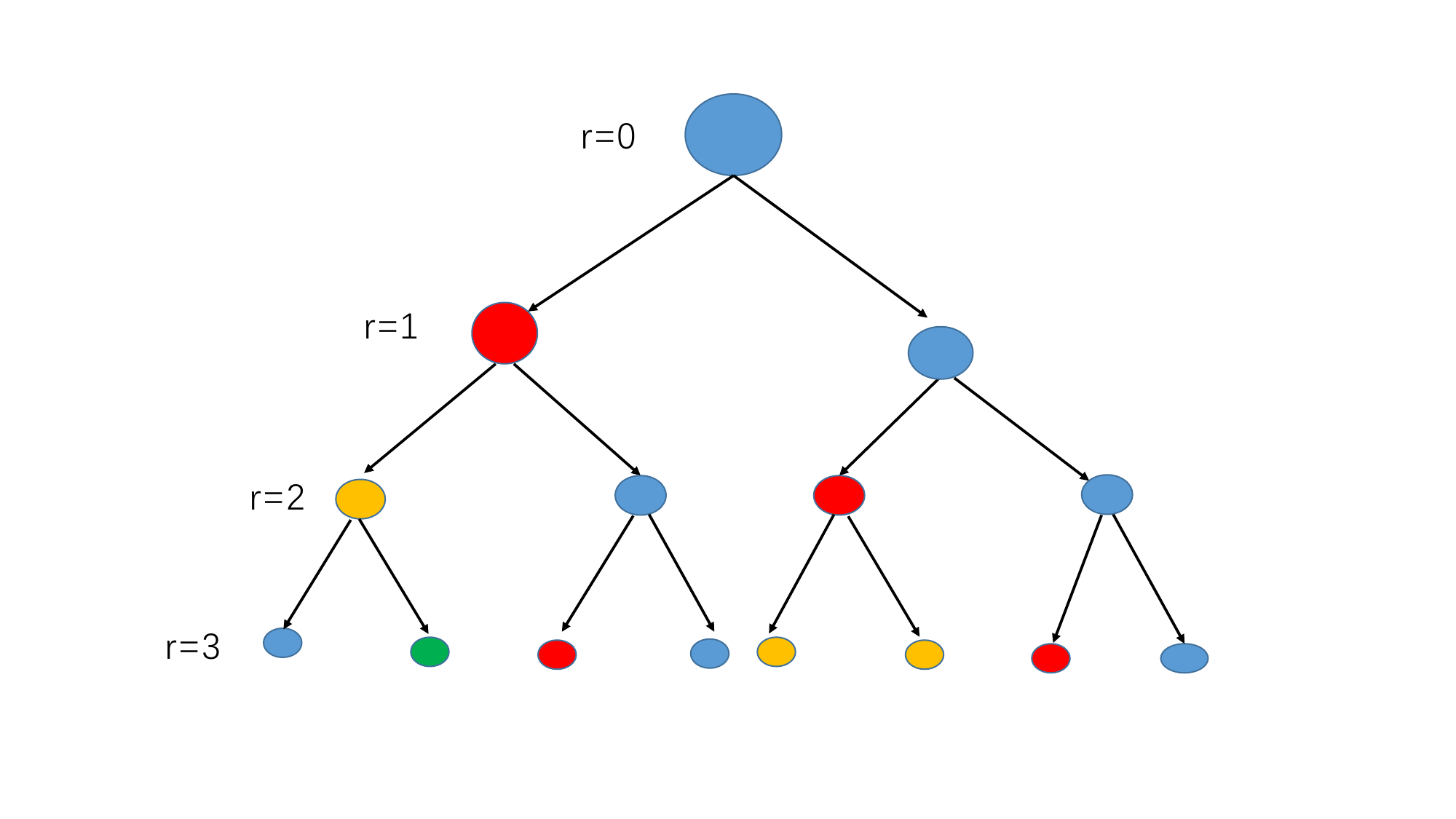}\\
 		(a)&(b)\\
 	\end{tabular}
 	\caption{The risk contagion description on two types of network structures. The red color denotes the nodes directly suffered from the outside cyber attacks, which are considered as origin contagion nodes. The nodes propagated by one-hop risk contagion are depicted in yellow and nodes propagated by two-hop contagion are denoted in green. }
 	\label{fig1}
 \end{figure}

In recent years, the modeling of risk contagion in network structure has attracted much attention from scholars. Accurate characterization of risk contagion not only provides guidance for risk managers, but also serves as a crucial foundation for cyber insurance pricing. 
The original one-hop risk propagation model was initially proposed by \cite{Kunonehop2003}, in which
a compromised node can propagate the risk to its direct neighbors and the risk does not propagate further than one-hop. Consequently, a compromised node is either caused by an external risk or its directly connected neighbors. However, in practical, an external incident could cause more than just one-hop depth due to the interconnectedness within the system.
 \cite{da_multivariate_2021} proposed a k-hop risk propagation model to
 describe the dynamics of node states, in which the depth of risk propagation can reach $k$ rounds rather than one round. Specifically, each external risk can propagate to its direct neighbors, and the infected neighbors continue propagating to their direct neighbors as well. 
 Figure \ref{fig1}(a) depicts a two-hop risk propagation scenario, where the surviving nodes are represented in blue, directly compromised nodes by external attacks are shown in red (referred to as origin contagion), nodes propagated by interconnected origin contagion are depicted in gold (representing one-hop propagation), and nodes propagated by two-hop propagation are denoted in green. An essential question arises regarding how to construct this type of k-hop risk propagation on a specific network structure. 
 Additionally, it is important to consider factors such as the location of origin contagion and node heterogeneity that may affect the probability of risk contagion.
 
 \subsection{Model description}
  In order to accurately describe the k-hop risk propagation,
  \cite{da_multivariate_2021} pointed out $k$ can be considered as the time scale (i.e., second, hour, or day). Therefore, their k-hop model describes the dynamic of risk propagation within the first $k$ unit times.
 In contrast, our risk propagation is limited to paths connecting a node with its descendants up to $k$ generation, which we refer to as path-based k-generation risk propagation. 
 In our risk model, a node is defined as an origin contagion if its risk loading is lower than the external risk size; this represents the initial step for k-generation risk propagation. There are several differences from existing models\cite{laszka_assessment_2018,da_multivariate_2021}. First, in our mechanism for risk propagation, "k-hop" means that an origin contagion node can successively propagate risks to its descendants until reaching $k$ generation, and this depth of risk propagation is referred to as "k-generation". Second, the size of risks decreases with propagation depth in our model. Finally, the impact of heterogeneity between nodes on risk propagation is also 
considered.

 Consider a system (entities, local network) consisting of $N$ nodes, which can be described as a tree-based graph $T_{R}=(V,E)$, where $V$ is the node set, $E$ is the edge set, and $R$ represents the radius of the tree-shaped network. 
 Assuming that the $T_{R}$ is usually rooted, and the tree is growing away from its root. Each node
 has branches leading to its descendants, and the branch number is denoted as $\rho$. 
 For the sake of completeness, we provide some basic concepts that are needed in our next work.
 A node $x$ is called an ancestor of $y$, and $y$ is a descendant of $x$, in short $x\in an_{G}(y)$ and $y\in de_{G}(x)$, if there exists a directed path from $x$ to $y$ in G. The nodes in $nd_{G}(x):=V\setminus(\{x\}\cup de_{G}(x))$ are called the nondescendants of $x$.
 In addition, for a sequence  $(w_{x})_{x\in V},$
 we also write $pa_{G}(w_{x})=(w_{y}:y\in pa_{G}(x))$ and define $an_{G}(w_{x})$ and $de_{G}(w_{x})$ analogously. 
 
 The network in Figure \ref{fig1}(b) represents a tree structure with a branch number of $\rho=2$, where each node has an equal branch size. It can be observed that the blue node at distance $r=o$ serves as the root node. At time $t_{1}$, the node at distance $r=1$ experiences an external risk attack and subsequently becomes an origin contagion. Consequently, the risk propagation initiates from this origin contagion to its first generation nodes, resulting in the compromise of the yellow node through propagation while others remain unaffected. Specifically, the yellow node is compromised by one-generation risk propagation, whereas the green node is compromised by two-generation risk propagation.
The node located at a distance of $r=2$ has been compromised by the second external attack at time $t_{2}$, and subsequently propagates two first-generation descendants. It is important to note that the occurrences of external attacks are independent. In this context, our main focus lies on analyzing the impact caused by each individual contagion node. 
 
To model the occurrence mechanism of the proposed path-based k-generation risk propagation, the adjustment coefficient $\beta_{k}$ is introduced to map the extent to which the risk magnitude changes with the depth of propagation; therefore, $\beta_{k}X$ represents the corresponding magnitude of risk when the external risk propagates $k$ generations on the network structure. For simplicity, nodes with the same radius away from the root have the same level of security, denoted as $c_{r+k}$.
In a regular tree-shaped network structure, each node has an identical number of descendants and predetermined paths leading to its k-generation descendants.
To facilitate the analysis, we consider a single path from the origin contagion node to one of its k-generation descendants.
 
We introduced a binary random variable $I_{r}^{(k)}$ to characterize the state of the event $\{\beta_{k}X>c_{r+k}\}$, 
which represents the occurrence of k-generation risk propagation from an origin contagion at distance $r$ away from the root.
 Note that a healthy node cannot propagate risk to its descendants, therefore, $\{I_{r}^{(k-1)}=0\}$ can not lead to the occurrence of $\{I_{r}^{(k)}=1\}$.
Thus the occurrence of event $\{\beta_{k}X>c_{r+k}\}$ is a conditional event, $k=1,2,\dots,k$, we have
\begin{equation}\label{form3.11}
\{\beta_{k}X>c_{r+k}\}=\{I_{r}^{(k)}=1\vert I_{r}^{(k-1)}=1\}.
\end{equation}

 Assume random variables $\bar{I}_{r}^{(k)},r=0,1,2,\dots, R$ represent the state of occurrence of the k-generation risk propagation on a path from an origin contagion to one of its k generation descendants.
 According to the aforementioned formula \eqref{form3.11},
  we can construct the following random event to express the occurrence of path-based k-generation risk contagion on a single path, denote $P_r{^{(k)}}$ the probability of the event $\{\bar{I}_{r}^{(k)}=1\}$, we have
 \begin{equation}\label{for8}
\begin{aligned}
P_{r}^{(k)}&=P(\bar{I}_{r}^{(k)}=1)=P(\bigcap_{l=1}^{k}\{I_{r}^{(l)}=1\})\\
 &=\prod_{l=1}^{k}P\{I_{r}^{(l)}=1\vert I_{r}^{(l-1)}=1\}P\{I_{r}^{(0)}=1\}.\\
&=\prod_{l=0}^{k}P\{ \beta_{l}X>c_{r+l}\}.
\end{aligned}
\end{equation}

To compute the explicit probability for $P(\bar{I}_{r}^{(k)}=1), k=1,2,\cdots$, assume that the risk size $X$ has the density function $f_{X}(x)$ and cumulative distribution $F_{X}(x)$, $\bar{H}_{X}(x)=1-F_{X}(x)$ represents the survival function,
\begin{equation}\label{for10}
    \bar{H}_{X}(x)=P(X>x),
\end{equation}
combining with \eqref{for8} and \eqref{for10}, the probability $P(\bar{I}_{r}^{(k)})$ could be obtained as follows
\begin{equation}\label{for11}
P(\bar{I}_{r}^{(k)}=1)=\prod_{l=0}^{k}\bar{H}(d_{l}),
\end{equation}
where $d_{l}=\frac{c_{r+l}}{\beta_{l}}$. It can be seen from the formula \eqref{for11}, the risk propagation depth $k$  and the location parameter $r$ are considered in the propagation probability. Unlike the risk contagion discussed in \cite{zhang_dynamic_2022}, the state of a node is mainly determined by three factors: external risk, recovery ability, and contagion from its direct neighbors.
Different from existing risk contagion mechanisms from the outside in, our proposed contagion model emphasizes the depth of risk propagation from the inside out and analyzes its consequences by considering the contagion along the entire path as a whole.

\begin{remark}
\begin{enumerate}
\item 
The probability that a node with a distance of $r$ is compromised by external risk attack can be computed by 
 \begin{equation*}
     P(\bar{I}_{r}^{(0)})=P(\beta_{0}X>c_{r})=\bar{H}(c_{r}).
 \end{equation*}
 \item
 The probability  $P(\bar{I}_{0}^{(1)}=1)=P\{I_{0}^{(l)}=1\vert I_{0}^{(0)}=1\}P\{I_{0}^{(0)}=1\}=\bar{H}(d_{1})\bar{H}(d_{0})$ represents an origin contagion root node propagates its direct offspring through one-hop risk propagation. 
\end{enumerate}
\end{remark}
Above, we give a discussion of the probability of generation-based k-hop risk propagation on a single path of the tree-shaped network structure. In addition, there is still a key problem, which is the severity of the k-generation risk propagation on a single path. Unlike the local loss size given in studies \cite{jevtic_dynamic_2020,da_multivariate_2021,zhang_dynamic_2022}, the losses for compromised nodes are assumed to be identical and independent.
We take a path with a depth of k as a unit, in order to reflect the impact of contagion depth k and security levels on the scale of local loss, it is convenient to use $\beta_{k}c_{r+k}X$ as the local loss on the path with k-generation risk contagion, denoted as
\begin{equation}\label{oneloss}
    Z_{r}^{(k)}=\beta_{k}c_{r+k}X.
\end{equation}
where parameter $c_{r+k}$ represents the impact coefficient on $\beta_{k}X$.
\begin{corollary}\label{loss1}
For k-generation risk propagation on a single path,
the mean and variance of the loss $Z_{r}^{(k)}$ can be easily derived using basic probabilistic properties
\begin{equation}\label{loss1mean}
    \begin{aligned}
 \mathbb{E}[Z_{r}^{(k)}]&=\beta_{k}c_{r+k}\mu_{X},\\
\mathbb{V}ar[Z_{r}^{(k)}]&=
\beta_{k}^{2}c_{r+k}^{2}\sigma_{X}^{2}.
    \end{aligned}
\end{equation}
\end{corollary}
In Corollary \ref{loss1mean}, to consider the impact of the location of origin contagion, the parameter $c_{r+k}$ is used to adjust the size of risk. Specifically, the loss size is $\beta_{k}c_{k}X$ when an origin contagion with a distance of $r=0$ propagates the risk to its $k$ generation descendants. 
 \subsection{Conditional independence}
 In the aforementioned context, we define the propagation probability on a path from an origin contagion to its $k$ generation nodes. However, there are $\rho^{k}$ paths for any node to its $\rho^{k}$ descendants of k generation. The interesting issue is how many paths to the $k$ generations are exposed to risk contagion caused by this origin contagion. 
 To solve this problem, the following $\rho^{k}$ dimensional joint probability distribution needs to be determined. 
 First, we introduce some representations to describe the paths of risk propagation in the whole network.
 For each node $x$ with a distance of $r$ away from the root which is compromised by external attacks, there is only one path to one of its k-generation descendant. Therefore, the collection of all paths to its k-generation descendants is denoted as follows
\begin{equation*}
   \Gamma_{r}^{(k)}(x)=\{x\to y\in E:d(x,y)=k\},
\end{equation*}
where $d(0,y)=r$ represents the risk propagation that starts at the root.
According to the characteristic of the branch structure of the tree-shaped network, the number of paths from node $x$ to its k generation descendants is $\rho^{k}$, that is
\begin{equation}
    \rho^{k}=card(\Gamma_{r}^{(k)}(x)).
\end{equation}

Based on our proposed risk contagion mechanism, not every path in the collection $\Gamma_{r}^{(k)}(x)$ succeeds in getting k-generation contagion caused by origin contagion. Whether each path is compromised has a certain probability of occurrence, which is related to the safety level of the node itself and the size of the risk.
 To solve the problem of how many paths in $\Gamma_{r}^{(k)}(x)$ suffer the k-generation risk contagion, we introduce the following $\rho^{k}$-dimensional Bernoulli random vector
 \begin{equation}\label{for6}
 ( \bar{I}_{1r}^{(k)},\dots, \bar{I}_{jr}^{(k)},\dots, \bar{I}_{\rho^{k} r}^{(k)}),
 \end{equation}
here, our focus is mainly on the probability properties of random vector 
 $( \bar{I}_{1r}^{(k)},\dots, \bar{I}_{jr}^{(k)},\dots, \bar{I}_{\rho^{k} r}^{(k)}).$ 
 The joint probability distribution should be given to compute the number of paths with occurrences of k-generation contagion.
 The classical approach for solving the joint probability distribution is the chain rule, which has the drawback of extensive computation in high dimensionality conditions.
 Additionally,
 for the random vector 
 $( \bar{I}_{1r}^{(k)},\dots, \bar{I}_{jr}^{(k)},\dots, \bar{I}_{\rho^{k} r}^{(k)})$, it is important to determine whether the occurrence of risk contagion on different paths depends on each other and how this relationship is affected by the states of other paths. 
 
In the subsequent context, the concept of d-separation in Bayesian network \cite{Graphical2009} is employed to address the aforementioned issue. Probability graphical models (PGMs) are widely utilized techniques that integrate probability theory and graph theory, primarily utilizing graphs to depict the probabilistic dependencies between variables, and have been successfully applied across various domains. Bayesian networks serve as a specific type of graphical model employed for representing variable dependencies. They are depicted by directed acyclic graphs (DAGs), where nodes symbolize variables and edges represent their interdependencies. Fortunately, within our study, we specifically focus on tree-shaped networks, 
which is a special direct acyclic graphs (DAGs). To solve the joint probability distribution for all nodes in DAGs, the concept of d-separation is a crucial tool to demonstrate the conditional independence among the nodes.
The following definition\cite{Graphical2009} is essential for the understanding of the Bayesian network. For a more detailed context, one can refer to \cite{Bayes1996}.
 
\begin{figure}[H]
 	\centering
 	\begin{tabular}{ccc}
 		\includegraphics[width=0.3\linewidth]{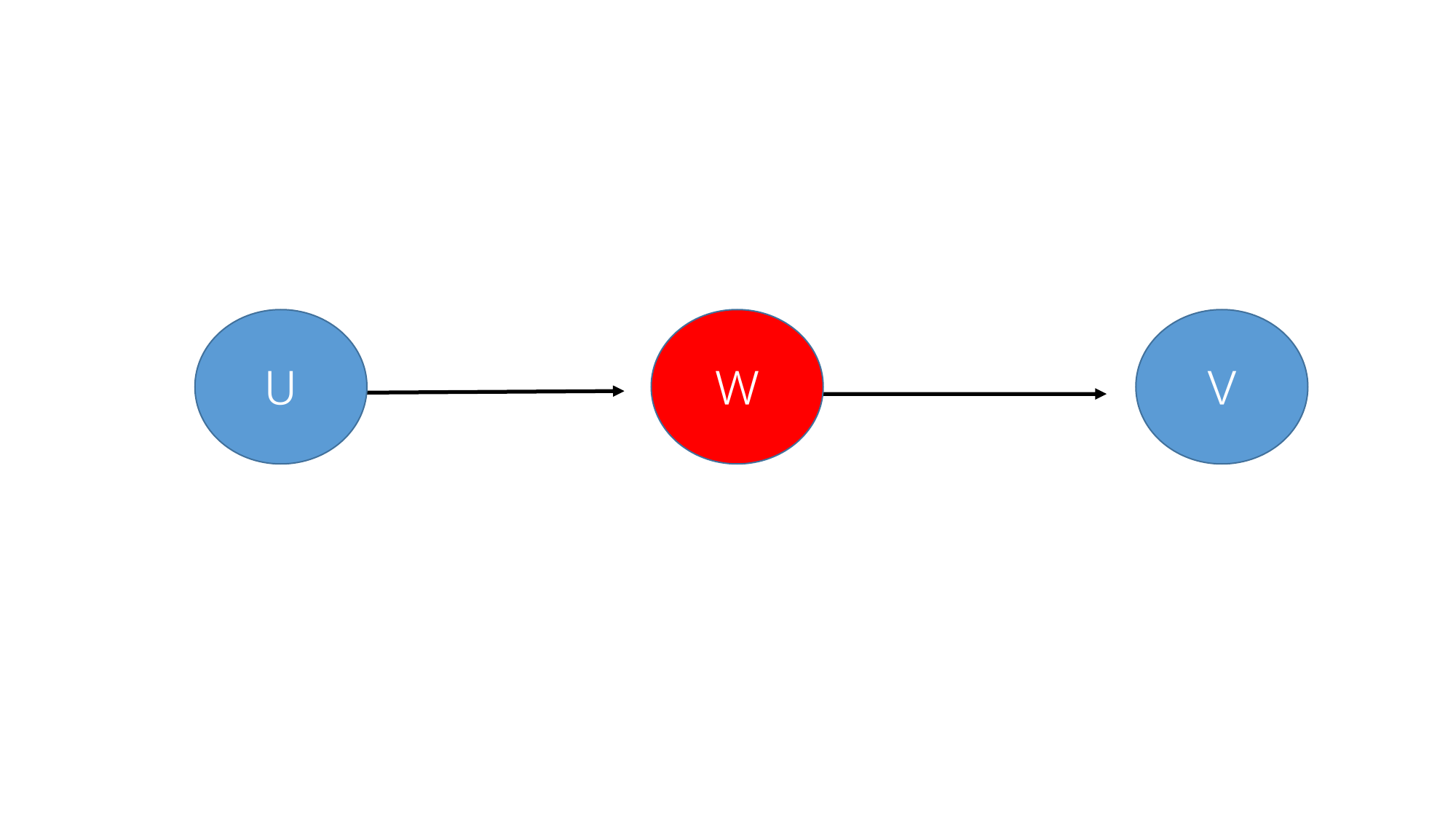}& 
 		\includegraphics[width=0.3\linewidth]{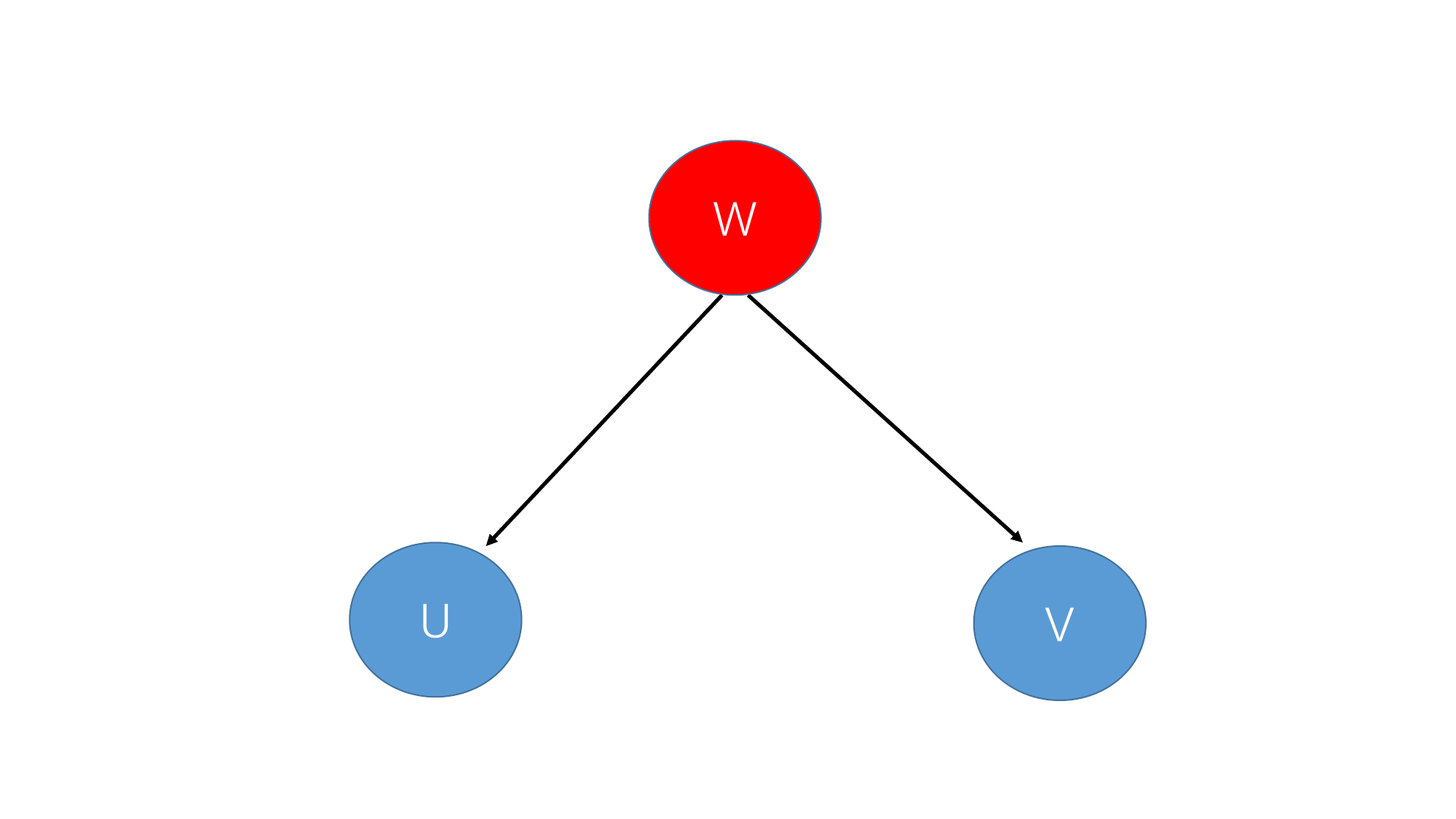}&
 		\includegraphics[width=0.3\linewidth]{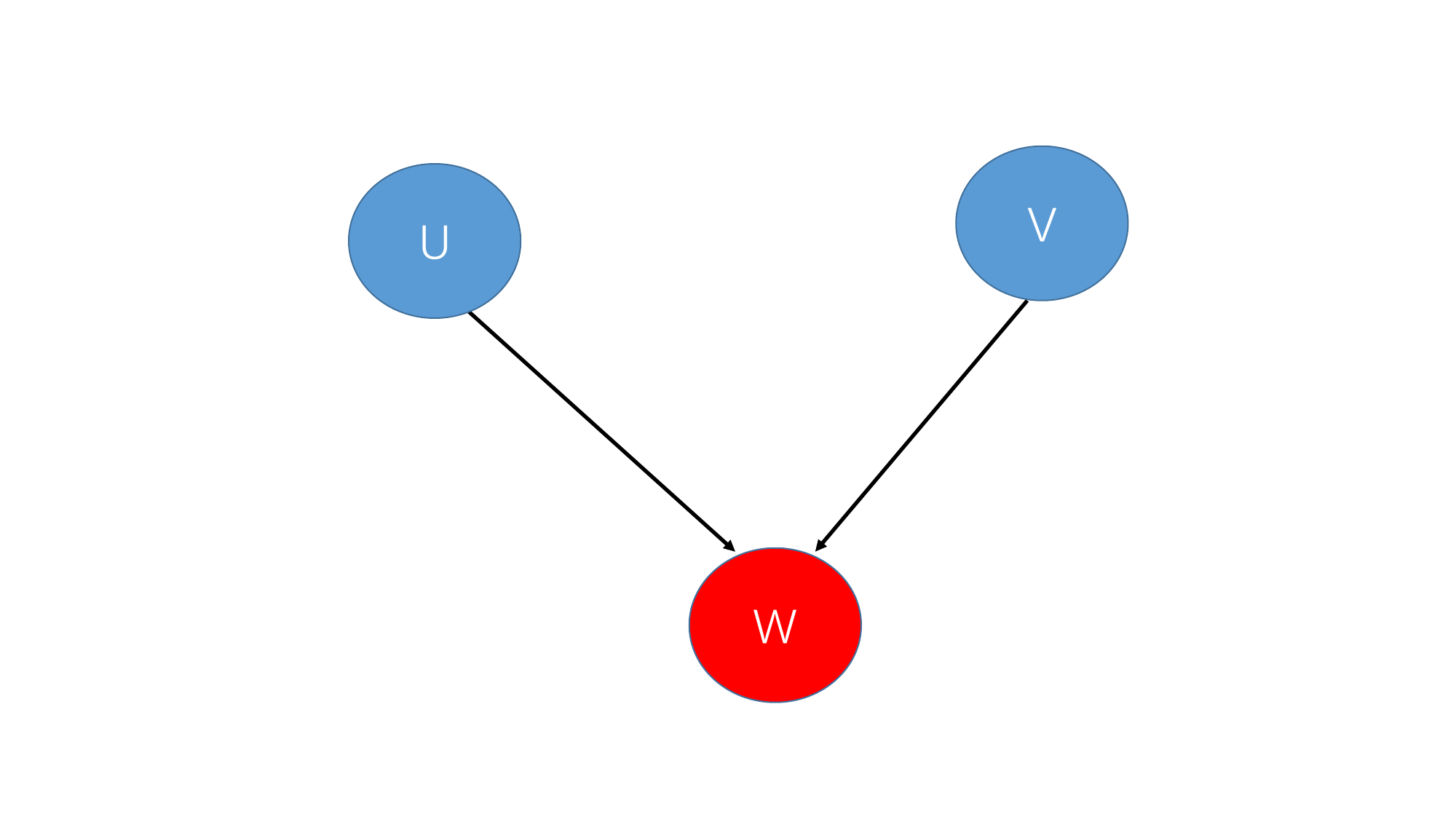}\\
 		(a) & (b)& (c)\\
 	\end{tabular}
 	\caption{Three common basic structures in directed graphs. }
 	\label{figthree}
 \end{figure}

 \begin{definition}\label{beyes}
 {\bf( Bayesian Network Factorization)}
  Given a DAG G=(V,E), a collection of 
  $\{W_{x}:x\in V\}$ of random variables taking values in a finite set $E$ is said to form a Bayesian network over G if for all $e=(e_{x}:x \in V)\in E^{\vert V\vert}$, there have
  \begin{equation}\label{bayes1}
     P[W_{v}=e]=\prod_{x\in V}P[W_{x}=e_{x}\vert pa_{G}(W_{x})=pa_{G}(e_{x})].
  \end{equation}
 \end{definition}
 An equivalent explanation for \eqref{bayes1} in \cite{Bayes1996},
 $\{W_{x}:x\in V \}$ forms a Bayesian network over G if and only if for every $x \in V$, the variable $W_{x}$ is conditionally independent of $W_{nd_{G}(x)}$ given $W_{pa_{G}(x)}$.
 In addition, the joint probability can be expressed as the product of several conditional probability distributions of each variable given its parents.
 we consider three basic Bayesian network (BN) structures for three variables and two arcs, which are given in Figure \ref{figthree}.  
The structures in Figure \ref{figthree} are called sequential, divergent, and convergent respectively. 
From the Figure \ref{figthree}, a crucial question is that given the state of variable $W$ in red, whether the states of other two variables $U$ and $V$ in blue are independent. 
To address this issue, an essential tool called d-separation\cite{Graphical2009},  which is a commonly used and effective criterion to determine whether a set $X$ of variables is independent of another set $Y$, given a third set $Z$. 
\begin{definition}\label{desp}
{\bf (d-separation)}
 Given a graph G=(V,E) and nodes $U$ and $V$ in $V$,
 for each trail between $U$ and $V$, the $U$ and $V$ are called d-separation, if the node $W$ in trail satisfy one of the following two conditions,
 \begin{enumerate}
     \item the connection of $W$ is serial or diverging and the state of $W$ is observed
     \item the connection of $W$ is concerging and neither the state of $W$ nor the state of any descendant of $W$ is observed.
 \end{enumerate}
\end{definition}
we give a simple example to get a better understanding of the use of d-separation in our risk contagion scenarios.
 \begin{example}\label{ex1}
 For the tree-shaped network structure with radius $R=2$ and the size of offspring $\rho=2$. Denote the node set $V=\{  V_{r},V_{r1},V_{r2},V_{r11},V_{r12},V_{r21},V_{r22}\}$, where $V_{r}$ represents the origin contagion at a distance $r$ from the root, $V_{rj},j=1,2$ represent j-th nodes of first generation, and 
 $V_{rjk},j=1,2;k=1,2$ represent the descendants of 2-generation of node $V_{r}$ 
\end{example}
 Assume that the node $V_{r}$ has been compromised by an external risk attack, we focus on the relationship between the state of its descendants. 
 The tree-shaped graph is comprised of the basic structure: sequential and divergent. By the definition of d-separation, we can easily obtain the following probability calculations.
\begin{equation}\label{exeq1}
\begin{aligned}
    P(V_{r1},V_{r2}\vert V_{r})&=\frac{P(V_{r},V_{r1},V_{r2})}{P(V_{r})}=\frac{P(V_{r1}\vert V_{r}),P(V_{r2}\vert V_{r})P(V_{r})}{P(V_{r})}\\
    &=P(V_{r1}\vert V_{r})P(V_{r2}\vert V_{r}),
    \end{aligned}
\end{equation}
the equation \eqref{exeq1} illustrates that the compromise states for descendants of an origin contagion are conditionally independently. The joint probability distribution can be expressed as the following,
\begin{equation*}
\begin{aligned}
    P(V_{r},V_{r1},V_{r2},V_{r11},V_{r12},V_{r21},V_{r22})&=P(V_{r})P(V_{r1},V_{r2}\vert V_{r})P(V_{r11},V_{r12}\vert V_{r1})P(V_{r21},V_{r22}\vert V_{r2})\\
    &=\prod_{j=1}^{2}\prod_{i=1}^{2}P(V_{rij}
    \vert V_{ri})P(V_{ri}\vert V_{r})P(V_{r}),\\
    \end{aligned}
\end{equation*}
compared with the results of chain rule, the number of parameters are decreased significantly. Furthermore, the question of joint probability distribution can be solved by independent conditional distribution, this greatly reduces the complexity of calculation. 
 The following Proposition \ref{prop2} proves that the occurrences of
 k-generation risk contagion on different paths with a common origin contagion are mutually independent and identical.
\begin{proposition}\label{prop2}
(The independence of k-generation contagion between multi-paths)  Given a original compromised node at distance $r$ away from root, the states of k-generation risk propagation on $\rho^{k}$ paths can be expressed as $(\bar{I}_{1r}^{(k)},\dots, \bar{I}_{r}^{(k)},\dots, \bar{I}_{\rho^{k} r}^{(k)})$, there have
\begin{equation}
 P(\bar{I}_{1r}^{(k)}=1,\dots,\bar{I}_{\rho^{k} r}^{(k)}=1)=\prod_{m=1}^{\rho^{k}}P(\bar{I}_{mr}^{(k)}=1).
\end{equation}
\begin{proof}
Here, we use mathematical induction to prove the mutual independence of the aforementioned events. Let $\bar{I}_{mr}^{(k)}, m=1,2,\dots,\rho^{k}$ 
represents the state of paths in which risk propagation from the original contagion at position $r$ to  its k-generation descendants.
According to the result of Corollary \ref{prop2},
given $k=1$, it has
\begin{equation}
P(\bigcap_{m=1}^{\rho}I_{mr}^{(1)}=1\vert \bar{I}_{r}^{(0)}=1)P(\bar{I}_{r}^{(0)}=1)=\prod_{m=1}^{\rho}P(I_{mr}^{(1)}=1\vert \bar{I}_{r}^{(0)}=1)P(\bar{I}_{r}^{(0)}=1),
\end{equation}

\begin{equation*}
  P(\bigcap_{m=1}^{\rho}\bar{I}_{mr}^{(1)}=1) =
  \prod_{m=1}^{\rho}P(\bar{I}_{mr}^{(1)}=1).
\end{equation*}
Assume that the above results hold for $k-1$, 
\begin{equation}\label{generation}
  P(\bigcap_{m=1}^{\rho^{k-1}}\bar{I}_{mr}^{(k-1)}=1) =
  \prod_{m=1}^{\rho^{k-1}}P(\bar{I}_{mr}^{(k-1)}=1),
\end{equation}
the equation \eqref{generation} indicates that for (k-1)-generation risk contagion caused by a origin contagion with location parameter $r$, the risk contagion among $\rho^{k-1}$ paths is mutually independently. For the location parameter $k$, we want to derive the following 
\begin{equation*}
  P(\bigcap_{m=1}^{\rho^{k}}\bar{I}_{mr}^{(k)}=1) 
  =\prod_{m=1}^{\rho^{k}}P(\bar{I}_{mr}^{(k)}=1).
\end{equation*}

For any fixed $m$ in $\{1,2,\dots,\rho^{k-1}\}$, there has $\rho$ paths towards its next offspring node, by the results of d-separation 
\begin{equation}\label{generationk}
\begin{aligned}
 P(\bigcap_{j=1}^{\rho}(\bar{I}_{jmr}^{(k)}=1))&=
 P(\bar{I}_{mr}^{(k-1)}=1,I_{1mr}^{(k)}=1,\dots,I_{\rho mr}^{(k)}=1)\\
 &=P(\bar{I}_{mr}^{(k-1)}=1)P(\bigcap_{j=1}^{\rho}I_{jmr}^{(k)}=1\vert \bar{I}_{mr}^{(k-1)}=1)\\
 &=P(\bar{I}_{mr}^{(k-1)}=1)\prod_{j=1}^{\rho}P(I_{jmr}^{(k)}=1\vert \bar{I}_{mr}^{(k-1)}=1)\\
 &=\prod_{j=1}^{\rho}P(\bar{I}_{jmr}^{(k)}=1),
 \end{aligned}
\end{equation}
where $j$ represents the next descendants of node $m$. Combining with the equation \eqref{generation}, and taking $m$ from 1 to $\rho^{k-1}$, we have
\begin{equation}\label{generation2}
\begin{aligned}
   P(\bigcap_{m=1}^{\rho^{k-1}}\bigcap_{j=1}^{\rho}(\bar{I}_{jmr}^{(k)}=1))
   &=\prod_{j=1}^{\rho}P(\bigcap_{m=1}^{\rho^{k-1}}(I_{jmr}^{(k)}=1,\bar{I}_{mr}^{(k-1)}=1))\\
   &=\prod_{j=1}^{\rho} \prod_{m=1}^{\rho^{k-1}}P(I_{jmr}^{(k)}=1, \bar{I}_{mr}^{(k-1)}=1)\\
   &=\prod_{j=1}^{\rho} \prod_{m=1}^{\rho^{k-1}}P(\bar{I}_{jmr}^{(k)}=1),\\
   \end{aligned}
\end{equation}
therefore, the \eqref{generationk} can be rewritten as 
\begin{equation}
  P(\bigcap_{m=1}^{\rho^{k}}\bar{I}_{mr}^{(k)}=1) =
  \prod_{m=1}^{\rho^{k}}P(\bar{I}_{mr}^{(k)}=1).
\end{equation}
In next, we give that the sequence $\{\bar{I}_{mr}^{(k)},m=1,2,\dots,\rho^{k}\}$ are identically distributed. For any $m \in \{1,2,\dots,\rho^{k}\}$, the $P(\bar{I}_{r}^{(k)}=1)$ is essentially a joint probability, namely 
\begin{equation*}
\begin{aligned}
    P(\bar{I}_{mr}^{(k)}=1)&=P(\beta_{k}X_{m}>c_{r+k}\vert \bar{I}_{mr}^{(k-1)}=1)P(\bar{I}_{mr}^{(k-1)}=1)\\
    &=P(\beta_{k}X_{m}>c_{r+k},\beta_{k-1}X_{m}>c_{r+k-1},\dots,\beta_{0}X_{m}>c_{r}),\\
    \end{aligned}
    \end{equation*}
the multivariate random sequences $(\beta_{k}X_{m},\beta_{k-1}X_{m},\dots,\beta_{0}X_{m}),m=1,2,\dots,\rho^{k}$ are identically distributed,
and the probability $P(\bar{I}_{mr}^{(k)}=1)$ is dependent on the distribution $F_{X}(x)$ and $\beta_{k}$.
Hence, the random variable sequence $\{\bar{I}_{mr}^{(k)}\}, m=1,2,3,\dots,\rho^{k}$ are mutually independently and identically distributed.   
\end{proof}
\end{proposition}
\begin{remark}
when $k=1$, there have 
\begin{equation*}
P(\bigcap_{m=1}^{\rho}I_{mjr}^{(1)}=1\vert I_{jr}^{(0)}=1)=\prod_{m=1}^{\rho}P(I_{mjr}^{(1)}=1\vert I_{jr}^{(0)}=1),
\end{equation*}
this illustrates the joint probability of state variables which caused by one round risk propagation originated from a node at a distance $r$ suffers from external risk attack.
In particular, when the external risk attack arrivals at the root node, 
\begin{equation*}
P(\bigcap_{m=1}^{\rho}I_{m0}^{(1)}=1\vert I_{0}^{(0)}=1)=\prod_{m=1}^{\rho}P(I_{m0}^{(1)}=1\vert I_{0}^{(0)}=1),
\end{equation*}
\end{remark}

An important conclusion can be drawn from the above analysis of the d-separation, that is, the occurrences of k-generation on multi-paths with a common origin contagion are independent and identically distributed.
 Therefore, for the random vector  
 $( \bar{I}_{1r}^{(k)},\dots, \bar{I}_{jr}^{(k)},\dots, \bar{I}_{\rho^{k} r}^{(k)})$, we can directly consider it as $\rho^{k}$-dimensional Bernoulli random variables that are mutually independently and identically. Finally, we can conclude the following results that are essential for the calculation of aggregate loss.

\begin{corollary}\label{number}
Denote $U_{r}^{(k)}$ the number of paths with the $k$ generations risk propagation, then we have
\begin{equation}\label{for9}
P\{ U_{r}^{(k)}=n\}=
\begin{pmatrix}
\rho^{k}\\n
\end{pmatrix}
[P_{r}^{(k)}]^{n}[1-P_{r}^{(k)}]^
{\rho^{k}-n},
\end{equation}
especially for the condition $n=\rho^{k}$, there have
$$P\{ U_{r}^{(k)}=\rho^{k}\}=[P(I_{r}^{(k)}=1)]^{\rho^{k}}.$$
\end{corollary}
By the properties of mean and variance, 
the $\mathbb{E}[U_{r}^{(k)}]$ and $\mathbb{V}ar[U_{r}^{(k)}]$ can be easily obtained as follow
\begin{equation}\label{formean}
\begin{aligned}
\mathbb{E}[U_{r}^{(k)}]&=\rho^{k}P_{r}^{(k)},\\
\mathbb{V}ar[U_{r}^{(k)}]&=\rho^{k}P_{r}^{(k)}(1-P_{r}^{(k)}).
\end{aligned}
\end{equation}
\subsection{The properties of aggregate loss }
A primary objective is to evaluate the risk propagation size of the entire network resulting from an origin contagion.
First, denote $S_{r}^{(k)}$ the risk size of the entire network that was caused by an origin contagion under path-based k-generation propagation.
Combining the results of Corollary \ref{loss1} and \ref{number}, we construct the local loss on the entire network under k-generation risk propagation 
\begin{equation}\label{newloss}
S_{r}^{(k)}=\sum\limits_{j=0}^{U_{r}^{(k)}}Z_{jr}^{(k)},
\end{equation}
where the number of paths with occurrences of k-generation risk propagation $U_{r}^{(k)}$ is a random variable and the corresponding loss size $Z_{r}^{jk}$ is given in Corollary \ref{loss1}. Note that the subscript in $Z_{jr}^{k}$ corresponds to the random number $U_{r}^{(k)}$, and $Z_{jr}^{(k)}$ has the same distribution as $Z_{r}^{(k)}$.
Based on the characteristics of the k-generation risk propagation, it can be found that $S_{r}^{(l)}$ is essentially a cumulative sum of local losses. This is different from the existing results, in which we do not add loss on all paths that originate from the origin contagion to its k generation descendants, but take it into account from the perspective of probability. Additionally, compared with the classical risk model \eqref{loss}, the risk frequency $U_{r}^{(k)}$ and the severity $Z_{jr}^{(k)}$ are  the generalizations of the $N_{t}$ and $Z_{i}$ in \eqref{loss}.
\begin{remark}
Note that, $Z_{jr}^{(k)}$ is defined on a single path from the origin contagion to one of its k-generation nodes. It can be concluded from \eqref{newloss} that our loss $S_{r}^{(k)}$ is equal to the actual loss when the depth of risk contagion $k=1$. 
However, with the parameter $k$ increasing,
the loss $S_{r}^{(k)}$ we construct is larger than the actual true loss because the overlapping loss should be subtracted when the risk propagates simultaneously on the same path for fewer than k-1 generations. Therefore, our loss model can be considered as provide an upper bound for the loss, which does not impact the subsequent parameter sensitivity analysis. 
\end{remark}
In general, the probability distribution of \eqref{newloss} is not easy to give, but we can derived its corresponding probability properties, which are crucial for the cyber risk insurance pricing.

\begin{theorem}\label{theo1}
For an origin contagion with a distance of r, the expectation and variance of the risk severity caused by the k-generation risk propagation on a given tree-shaped network structure are obtained by the following:
\begin{equation}
\begin{aligned}
 \mathbb{E}[S_{r}^{(k)}]&=\mathbb{E}[\sum_{j=0}^{U_{r}^{(k)}}Z_{jr}^{(k)}]=\mathbb{E}[U_{r}^{(k)}]\mathbb{E}[Z_{jr}^{(k)}]=
 \rho^{k}P_{r}^{(k)}\beta_{k}c_{r+k}\mu_{X},
 \\
    \mathbb{V}ar[S_{k}^{(k)}]&=
    \mathbb{V}ar[\sum_{j=0}^{U_{r}^{(k)}}Z_{jr}^{(k)}]\\
&=\mathbb{E}[U_{r}^{(k)}]\mathbb{V}ar[Z_{r}^{(k)}]+\mathbb{V}ar[U_{r}^{(k)}](\mathbb{E}[Z_{r}^{(k)})^{2}\\
&=\rho^{k}P_{r}^{(k)}\beta_{k}^{2}c_{r+k}^{2}\sigma_{X}^{2}+\rho_{k}P_{r}^{k}(1-P_{r}^{(k)})\beta_{k}^{2}c_{r+k}^{2}\mu_{X}^{2}     \\
&=\rho^{k}P_{r}^{(k)}\beta_{k}^{2}c_{r+k}^{2}(\sigma_{X}^{2}+(1-P_{r}^{(k)})\mu_{X}^{2})
    \end{aligned}
\end{equation}
\end{theorem}
To derive the probabilistic properties of the aggregate loss \eqref{form2} under the proposed path-based k-generation contagion model, let go back to the collective risk model given in Subsection \ref{collective}.
Denote $L_{rt}^{(k)}$ the aggregate loss on a tree-shaped network until the time $t$, in which the parameters $r$ and $k$ characterize the type of risk propagation.
The mean and variance of local loss are derived in Theorem \ref{theo1}. For the frequency of the external risk arrivals, the marked Poisson process with intensity $\mu$ is used. 
Substituting the results of Theorem \ref{theo1} into the equation \eqref{form2}, we can derive the following explicit mean and variance of aggregate loss which are crucial for the insurance pricing.
\begin{theorem}\label{theo2}
 The mean and variance of aggregate loss based on  the k-generation risk propagation can be derived as follows
 \begin{equation}
 \begin{aligned}
     \mathbb{E}[L_{rt}^{(k)}]&=\mu \rho^{k}P_{r}^{(k)}\beta_{k}c_{k+l}\mu_{X} t,\\
     \mathbb{V}ar[L_{rt}^{(k)}]&=
    \mu t\beta_{k}^{2}c_{r+k}^{2}[\rho^{k}P_{r}^{(k)}\sigma_{X}^{2}+\rho^{k}P_{r}^{k}(1-P_{r}^{(k)})\mu_{X}^{2}]+\mu t
    (\rho^{k}P_{r}^{(k)}\beta_{k}c_{r+k}\mu_{X})^{2},
    \end{aligned}
 \end{equation}
 where $\mathbb{E}[S_{r}^{(k)}]$ and $\mathbb{V}ar[S_{r}^{(k)}]$ are given in Theorem \ref{theo1}, and $P_{r}^{(k)}=\prod\limits_{l=0}^{k}\bar{H}(d_{l})$.
 \end{theorem}
It can be concluded that the mean of aggregate loss is dependent with the depth of risk contagion, the location of origin contagion, and the time $t$. Compared with existing works, the proposed model is more flexible to deal with the impact of essential factors. In particular, when the risk contagion originates from the root node, the following mean and variance can be derived.
\begin{corollary}
When the risk contagion starts from the root node,
there have
\begin{equation}
 \begin{aligned}
     \mathbb{E}[L_{0t}^{(k)}]&=\mu \rho^{k}P_{0}^{(k)}\beta_{k}c_{k}\mu_{X} t,\\
     \mathbb{V}ar[L_{0t}^{(k)}]&=
    \mu t \beta_{k}^{2}c_{k}^{2}[\rho^{k}P_{0}^{(k)}\sigma_{X}^{2}+\rho^{k}P_{0}^{k}(1-P_{0}^{(k)})\mu_{X}^{2}]+\mu t
    (\rho^{k}P_{0}^{(k)}\beta_{k}c_{k}\mu_{X})^{2},
    \end{aligned}
 \end{equation}
where$P_{0}^{(k)}=\prod\limits_{l=0}^{k}\bar{H}(d_{l}),d_{l}=\frac{c_{l}}{\beta_{l}}.$
\end{corollary}

\section{Numerical application}\label{sec4}
To get a better understanding of the proposed risk propagation mechanisms and characteristics,
in this section, we conduct sensitivity analysis to validate the impact of parameters on the mean, variance of propagation size $U_{r}^{(k)}$ and local loss $S_{r}^{(k)}$, specifically giving an application of cyber pricing under two kinds of pricing principles. First, we make the following assumption:
\begin{enumerate}
    \item the arrival times $\{t_{i},(i=1,2,\dots,N_{t})\}$ of external risk follow a marked homogeneous Poisson process with intensity $\mu=1.5$, and each $t_{i}$ equipped with a marked stochastic tree-shaped network $T_{R}^{i}$ and an external risk size $X$, that is, two components of the mark are employed jointly to determine the probability of events in which nodes suffer from loss from a given external attack event.    
    \item the external risk size $X$ follows the Gamma distribution with parameters $(\alpha,\lambda)=(5,1)$. The risk size adjust coefficient $\beta_{l}=0.95^{l}, l=0,1,2,\dots,k$ and $\beta_{0}=1$.
    
    \item The stochastic tree-shaped networks $T_{R}^{i}$
    are homogeneous with $R=30$ and the size of descendants $\rho=2$, in which the security levels of nodes vary with the location parameter $r$.
     The security levels $c_{r}$ that located at a radius $r$ away from the root are sampled from $C_{r}U(0,1).$
\end{enumerate}
\subsection{Cyber risk propagation}
\begin{figure}[H]
 	\centering
 	\begin{tabular}{cc}
 		\includegraphics[width=0.46\linewidth]{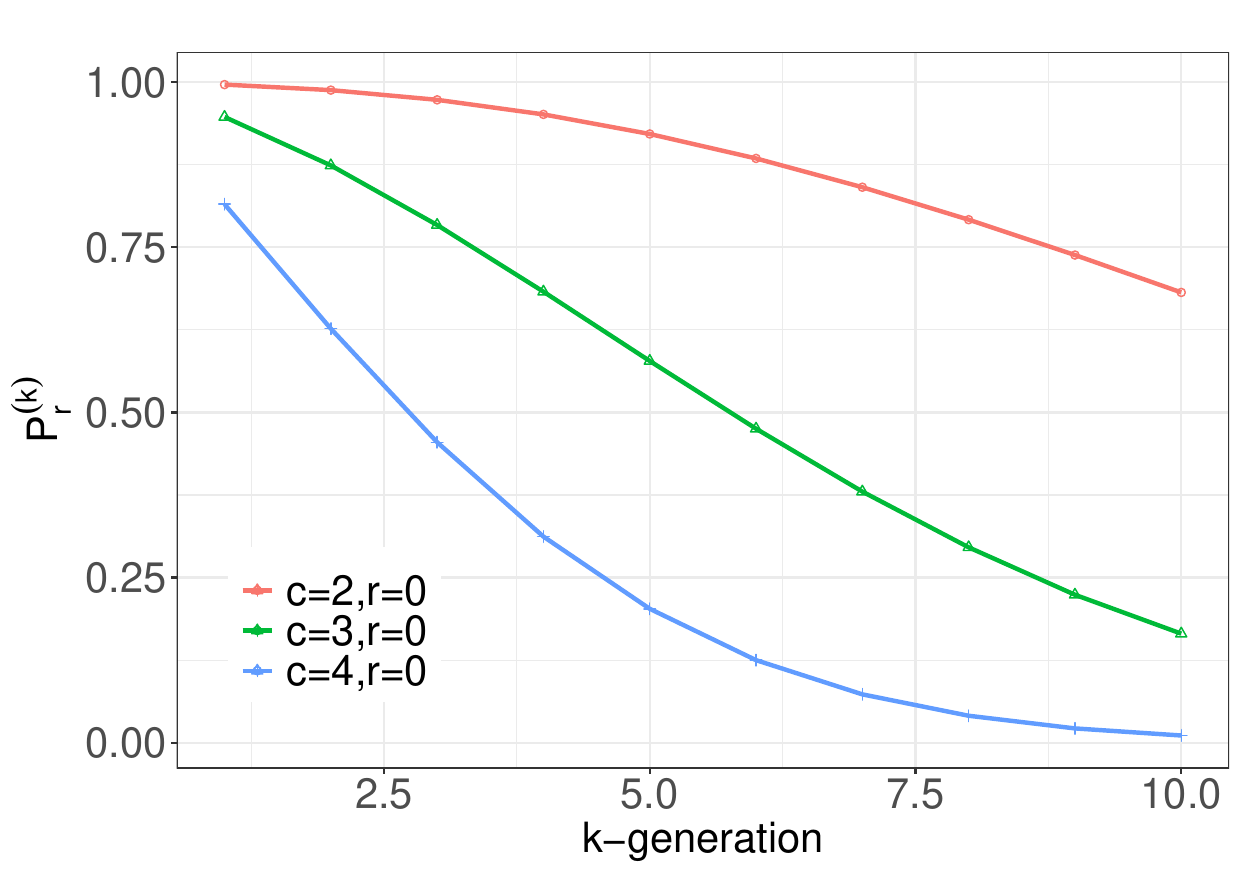}& 
 		\includegraphics[width=0.46\linewidth]{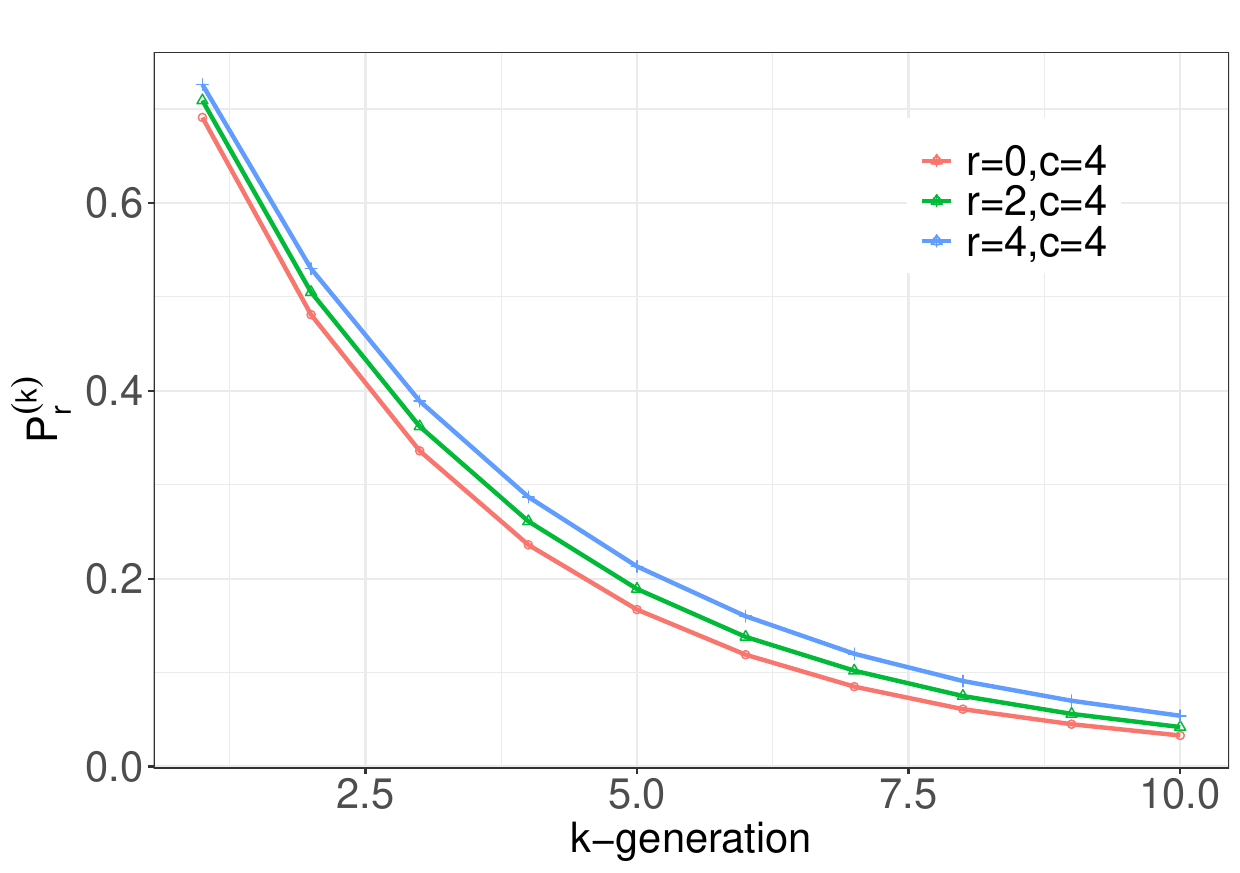}\\
 		(a)&(b)\\
 	\end{tabular}
 	\caption{Analysis of the influence of node security level and original risk location on risk propagation probability.}
 	\label{comprob}
 \end{figure}

 We compute the $P_{r}^{(k)}$ for the network described in above assumption. Combining with the formulas given in \eqref{for11}, the probability $P_{r}^{(k)}$ has the following explicit form
\begin{equation*}
    P_{r}^{(k)}=\prod_{l=0}^{k}\bar{H}(d_{l})=\prod_{l=0}^{k}(1-F_{X}(\frac{c_{r+l}}{\beta_{l}})).
\end{equation*}
Figure \ref{comprob} illustrates the impact of location parameter $r$ and security level $c$ on the probability of the k-generation risk propagation. Figure \ref{comprob}(a) shows that $P_{r}^{(k)}$ decreases as depth $k$ increases, when the origin attack propagates from the location $r=0$. 
The lines with three colors indicate that, given the location of origin propagation, the security level of nodes has significantly affects
the probability of path-based k-generation propagation. Specifically, enhancing the security level of nodes can effectively reduce the probability of risk propagation. Figure \ref{comprob}(b) shows that for a given security level of $c=4$, the probability $P_{r}^{(k)}$ varies with different locations of origin contagion. The further away the original contagion is from the root node, the higher the contagion probability is. This is mainly due to better protection and higher security levels typically found in root nodes 
compared to their descendant node's lower risk defense abilities.

For each origin contagion node, the number of paths to its $k$ generation nodes is $\rho^{k}$. Therefore, the focus is on the expected number of paths caused by k-generation risk contagion. 
Combining with the results of Corollary \ref{number},
the expected number of paths with the occurrence of k-generation risk contagion is $\rho^{k}P_{r}^{(k)}$.
From Table \ref{tabnum1}, we can conclude that the number of compromised paths varies with different origin contagion locations.
The further the origin contagion node away from the root,
the expected number of compromised paths is larger.
\begin{table}[htp]
\centering
\caption{\small The expected number of paths with the occurrence of k-generation risk contagion under the different locations of origin contagion.}
 \label{tabnum1}
 \small
\begin{tabular}{ccccccccccc}
\toprule 
k  & 1  & 2 & 3 & 4 & 5 & 6 & 7 & 8 & 9 & 10\\
Paths & 2 &  4 & 8 &16 & 32 & 64 & 128  & 256 & 512 & 1024\\

$\mathbb{E}[U_{0}^{(k)}]$& 1.3 & 1.6 & 2.1 & 2.9& 4.0 & 5.6 & 8.1 & 12.5 & 18.1 &29 \\
$\mathbb{E}[U_{2}^{(k)}]$&1.4  &1.9  & 2.8 & 4.1 & 6.1& 9.4 & 14.8 & 23.9& 39.5& 66.7\\
$\mathbb{E}[U_{4}^{(k)}]$&1.5 & 2.3 &3.5  &5.6 & 9.0 & 14.7 &25.0 & 43.3& 76.1& 136.5\\
\bottomrule 
\end{tabular}
\end{table}

To get a better sensitivity analysis of node security heterogeneity on the results of risk contagion, we are also interested in the effect of improving the security level of nodes at a given origin contagion location on the number of risk contagion.
Table \ref{tabnum2} shows that given an origin contagion location $r=0$, the number of k-generation contagion paths is decreasing dramatically with the improvement of security level from $c=2$ to $c=4$. Therefore, for nodes (enterprises) within an interconnected structural system, enhancing their own security level is an effective way to reduce the probability of being compromised.
\begin{table}[htp]
\centering
\caption{\small The expected number of paths with the occurrence of k-generation risk contagion under the different levels of security.}
 \label{tabnum2}
 \small
\begin{tabular}{ccccccccccc}
\toprule 
k  & 1  & 2 & 3 & 4 & 5 & 6 & 7 & 8 & 9 & 10\\
Paths & 2 &  4 & 8 &16 & 32 & 64 & 128  & 256 & 512 & 1024\\
$\mathbb{E}[U_{0}^{(k)}]\vert c=2$& 2.0 & 4.0& 7.8 & 15.2& 29.5 & 56.6& 107.6 & 202.7 & 378.0 &698.1\\
$\mathbb{E}[U_{0}^{(k)}]\vert c=3$&1.9  &3.5 & 6.3 &10.9 & 18.5 & 30.4 & 48.7 &75.7& 114.7&169.1 \\
$\mathbb{E}[U_{0}^{(k)}]\vert c=4$& 1.6 &2.5 & 3.6 &5.0 &6.5 & 8.0 & 9.4 &10.5 & 11.1& 11.3\\
\bottomrule 
\end{tabular}
\end{table}

\begin{figure}[H]
 	\centering
 	\begin{tabular}{cc}
 		\includegraphics[width=0.46\linewidth]{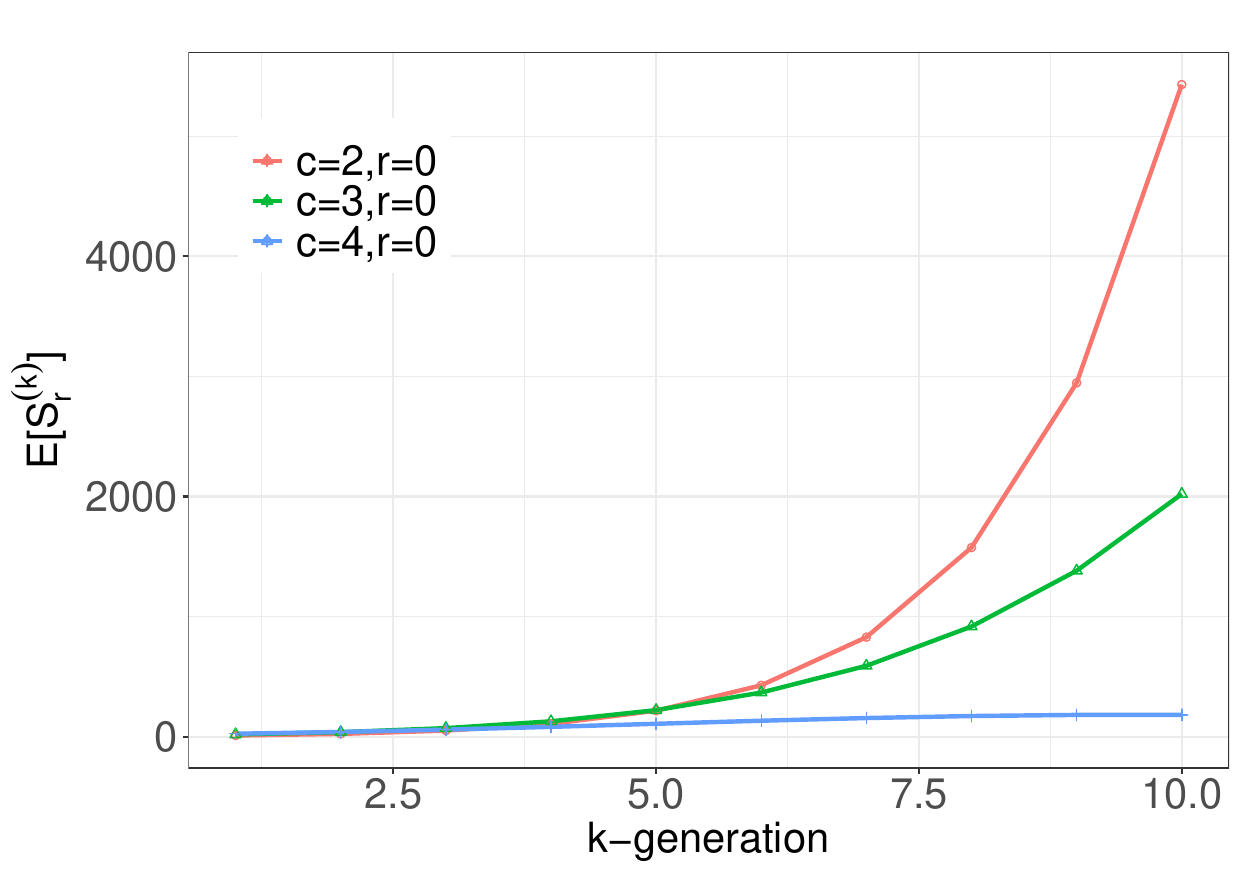}& 
 		\includegraphics[width=0.46\linewidth]{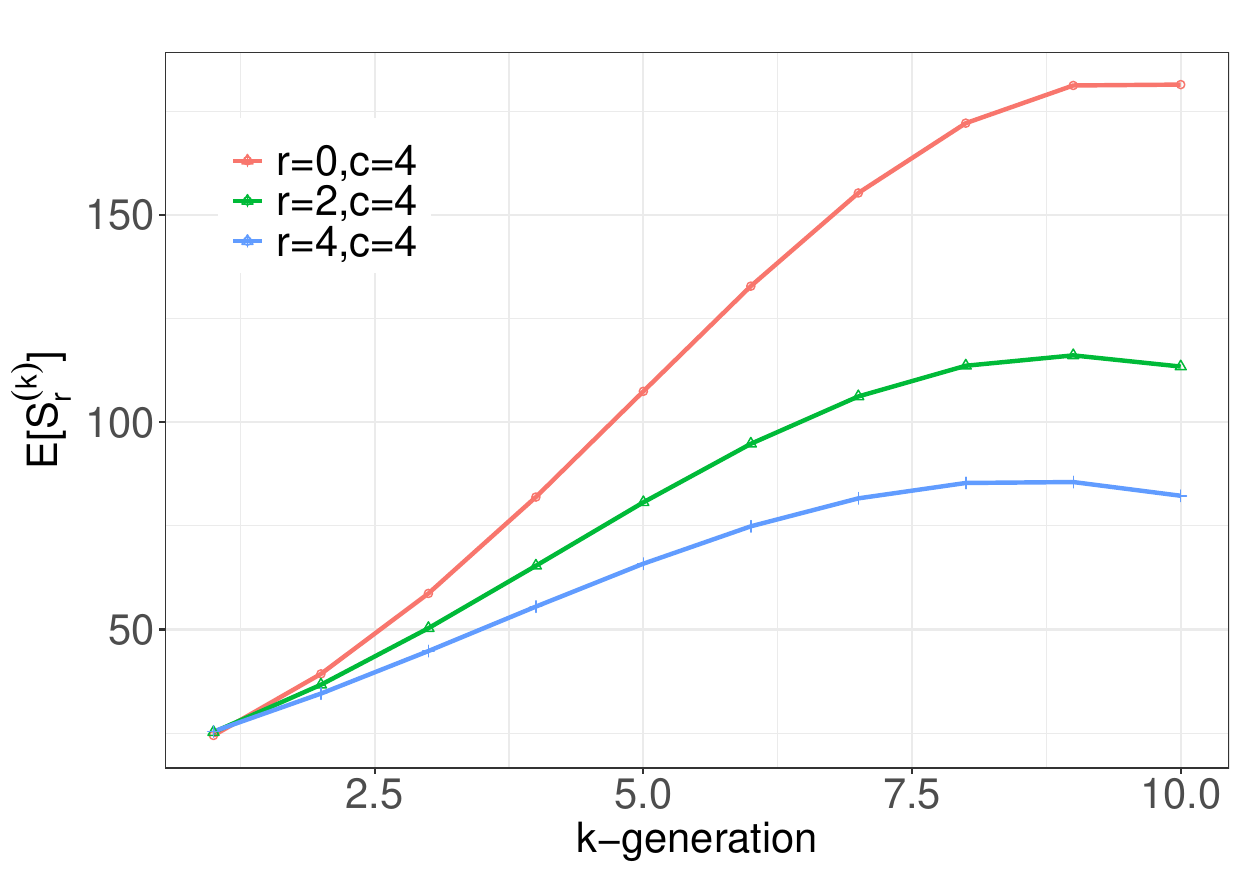}\\
 		(a)&(b)\\
 	\end{tabular}
 	\caption{The expected local loss caused by the occurrence of k-generation risk contagion on a single path.}
 	\label{localloss}
 \end{figure}

The $S_{r}^{(k)}$ gives the local loss caused by the origin contagion, which is dependent not only on the location parameter $k$  but also the security levels of nodes in the paths. Assume in scenario (a) that given the origin contagion location $r=0$, the risk contagion starts at the root node to its descendants.
In Figure \ref{localloss}(a), the red line indicates that as the depth of contagion increases, the local loss increases sharply. Under the three different origin contagion locations with varying security levels, there is a significant difference in the local loss caused by the k-generation risk contagion. At the origin contagion location with a higher security level, the size of $S_{r}^{(k)}$ is small, and the increase is stable.
Under the second scenario (b), the size of  $S_{r}^{(k)}$ is decreasing with the origin contagion parameter $r$ from 0 to 4, which means the further origin contagion location has an intensive impact compared with the location close to root.
Although the root node has a smaller probability of external risk attacks than its descendants, the root node generally plays a more important role in the network system, and the economic loss caused by risk attacks is often greater than that of ordinary nodes. Therefore, for insurers, it is still necessary to consider the risk control of the root node.

In summary, the aforementioned numerical study yields the following useful conclusions:
\begin{enumerate}
    \item the location of origin contagion plays a pivotal role in influencing risk contagion.
    \item the numerical results of k-generation contagion probability, compromise size and local loss can all demonstrate the significant impact of security level heterogeneity on risk contagion. 
\end{enumerate}

\subsection{Pricing cyber risk}
Based on the explicit mean and variance formulas of the aggregate loss for a series of random tree-shaped networks given in Theorem \ref{theo2}, in this subsection, we conducted the numeric calculation of cyber insurance pricing under two commonly used pricing principles
\begin{enumerate}
  \item 
the actuarial fair premium principle: $\mathbb{E}[L]$ 
  \item
the standard deviation principle: 
$\mathbb{E}[L]+\delta\sqrt{\mathbb{V}ar[L]}$.
\end{enumerate}
For more detailed information on premium principles, one can refer to \cite{modern2008}. Following the commonly
used configuration, we maintain the parameter $\delta=0.1$.
It is assumed that the external risk size $X_{i}$ follows a Gamma distribution $Ga(5,1)$, which is frequently used for modeling  risk severity due to its ability to capture the heavy-tailed characteristic existing in risk losses. Additionally, the Normal distribution with parameter $(\mu,\sigma^{2})=(5,4)$ is employed to compare the impact of loss distribution choices of loss distribution on pricing outcomes.
The selection of two distinct risk size distributions in our experiments aims solely at facilitating a clearer contrast in parameter results. In practical applications, 
the selection of risk size distribution is based on accurate estimation from a large amount of historical claim data, and this issue is beyond the scope of our discussion. Another essential parameter is the intensity $\mu$ of external risk arrival sequences, where $\mu=1.5$ is adopted here. 
To assess how heterogeneity in security levels and origin risk contagion location affects premiums, we assume three origin contagion locations described by $r = 0, r = 2$, and $r = 4$ respectively. This essentially characterizes how different levels impact premiums since security levels vary with parameter $r$.

There are several findings that can be concluded from the numerical results presented in Table \ref{tab1}. Firstly, for a fixed origin contagion location $r$, the premium under each principle and risk distribution consistently exhibits lower values for the security level $c=4$ compared to $c=2$. 
Hence, it is imperative to conduct an appropriate external security audit or employ self-reporting methods beforehand in order to mitigate information asymmetry\cite{yang_optimal_2019} and accurately estimate premiums. 
Secondly, when considering identical levels of security and risk size, the standard deviation principle yields slightly higher premium outcomes compared to the expectation premium principle. 
\begin{table}[htp]
\centering
\caption{\small The impact of nodes with different risk loading thresholds on risk pricing outcomes is considered under two pricing principles and two risk scale distribution. }
 \label{tab1}
 \small
 \begin{tabular}[h]{lcccccccccccc}
 \hline
 \multicolumn{5}{r}{$\mathbb{E}[L_{t}]$} & & & &\multicolumn{5}{r}{$\mathbb{E}[L_{t}]+\delta\sqrt{\mathbb{V}ar[L_{t}]}$ }\\
 \cline{2-6}\cline{9-13}
        &\multicolumn{2}{c}{$X\sim Ga(5,1)$} & & \multicolumn{2}{c}{$X\sim N(5,4)$}& &&\multicolumn{2}{c}{$X\sim Ga(5,1)$}& & \multicolumn{2}{c}{$X\sim N(5,4)$}\\
   \cline{2-3}\cline{5-6}\cline{9-10}\cline{12-13}
            k   & {$c=2$}  & {$c=4$}  &  & {$c=2$} &  {$c=4$}  &  &  &  {$c=2$} &  {$c=4$}  & & {$c=2$}   &  {$c=4$}    \\
 \hline
1& 149 & 367 &  & 126 & 311 &  &  & 155 & 385  & & 139 & 336   \\
2& 349 & 590 &  & 246 & 434 &  &  & 357 & 622 & & 267 & 470  \\
3 & 763 & 881 &  &  445 & 578 &  &  & 777 & 927  & & 478 & 623 \\
 4& 1599 & 1230&  &  766 & 737 &  &  & 1623 & 1291  & &  814& 792   \\
 5 & 3244 & 1612 &  & 1274 & 905 &  &  & 3284 & 1687  & & 1339 & 968    \\
 6 & 6418 & 1993 &  &  2059 & 1076 &  &  & 6483 & 2080 & & 2146 & 1146 \\
7 & 12429& 2329 &  &  3256 & 1239 &  &  & 12530 & 2426 & &  3369 & 1315   \\
8 & 23628 & 2582 &  & 5053 & 1386 &  &  & 23785 & 2686& & 5198 & 1467   \\
 9 & 44194 & 2718 &  & 7720 & 1509 &  &  & 44429 & 2826 & & 7902 & 1594   \\
10 & 81445 & 2722 &  & 11635 & 1600 &  &  & 81794& 2830  & & 11862 & 1688    \\
\hline
 \end{tabular}
 \end{table}
 
The results presented in Table \ref{tab2} highlight the importance of the origin contagion location for premium outcomes at a fixed security level. It is evident from the table that premiums linked to an origin contagion location r = 0 consistently exhibit lower values when compared to those associated with an origin contagion location r = 4.
Based on the results of the k-generation risk contagion provided in Theorem \ref{theo1}, it can be understood that for non-root nodes with lower security levels, vigilance in their security protection cannot be relaxed. This is because once they are attacked, the probability of risk contagion to similar nodes is high, which increases the number of compromised nodes and ultimately leads to higher premiums.

\begin{table}[htp]
\centering
\caption{\small The impact of origin contagions with different locations on risk pricing outcomes is considered under two pricing principles and two risk scale distribution.}
 \label{tab2}
 \small
 \begin{tabular}[h]{lcccccccccccc}
 \hline
 \multicolumn{5}{r}{$\mathbb{E}[L_{t}]$} & & & &\multicolumn{5}{r}{$\mathbb{E}[L_{t}]+\delta\sqrt{\mathbb{V}ar[L_{t}]}$ }\\
 \cline{2-6}\cline{9-13}
        &\multicolumn{2}{c}{$X\sim Ga(5,1)$} & & \multicolumn{2}{c}{$X\sim N(5,4)$}& &&\multicolumn{2}{c}{$X\sim Ga(5,1)$}& & \multicolumn{2}{c}{$X\sim N(5,4)$}\\
   \cline{2-3}\cline{5-6}\cline{9-10}\cline{12-13}
            k   & {$r=0$}  & {$r=4$}  &  & {$r=0$} &  {$r=4$}  &  &  &  {$r=0$} &  {$r=4$}  & & {$r=0$}   &  {$r=4$}    \\
 \hline
1 & 367 & 381 &  &  311 & 343&  &  & 385 & 404 & &  336 & 371 \\
2& 590 & 519 &  & 434 & 429&  &  & 622 & 553  & & 470& 466\\
 3 & 881 & 672 &  &  578 & 521 &  &  & 927 & 717  & & 623 & 566  \\
 4& 1230 & 833 &  &  737 & 617 &  &  & 1291 & 887  & &  792 & 669\\
 5 & 1612 & 988 &  & 905 & 714 &  &  & 1687 & 1050 & & 968 & 771 \\
 6 & 1993 & 1124 &  &  1076 & 807 &  &  & 2080 &  1191& & 1146 & 869\\
7 & 2329 & 1225 &  &  1239 & 893 &  &  & 2426 & 1296  & &  1315 & 958 \\
8 & 2582 & 1280 &  & 1386 & 966 &  &  & 2686& 1354  & & 1467 & 1033 \\
 9 & 2718 & 1284 &  & 1509 & 1022 &  &  & 2826 & 1358 & & 1594 & 1092 \\
 10 & 2722 & 1234 &  & 1600 & 1059&  &  & 2830 & 1307  & & 1688 & 1130 \\
\hline
 \end{tabular}
 \end{table}

\section{Conclusion}\label{sec5}
This work focus on the modeling of cyber risk propagation and aggregate loss in network with tree-shaped topologies. We propose a kind of path-based k-generation risk contagion model for tree-shaped network structures, in which the impact of the heterogeneous of nodes security levels and the location of origin contagion are incorporated. The properties of conditional independence is derived using the concept of d-separate in Bayesian network and the number of local propagation nodes is calculated in a closed form. We further derived the explicit expressions of expectation and variance which are essential for the pricing of cyber insurance. To get a better understanding of the proposed risk contagion model, we conduct the numerical calculation to examine the impact of location parameter and security level on contagion probability and aggregate loss. Several useful findings are concluded which are of great value for cyber risk managers and insurers.

Expanding on related work, we have constructed a risk contagion mechanism based on probabilistic distribution, rather than representing contagion probabilities with constants, which greatly enhances the interpretability of the risk contagion model. However, the scenarios in which actual risks occur are much more complex. Based on the work presented in this paper, there are still several considerations to take into account, such as discussing the proposed contagion mechanism on more general network topologies. Additionally, our work is proposed within the framework of mathematical models. To enhance the feasibility of the model, more detailed industry-specific background information should be taken into account, such as \cite{Lanchier_2023}. 
\vskip 0.5cm
\noindent
{\bf Acknowledgments}

 {\small This work was supported by National Natural Science Foundation of China (12361029,12161050).}
{\bf Declaration of Competing Interest}

The authors have no conflicts of interest to declare.
\bibliographystyle{unsrt}
\bibliography{bib}

\end{document}